\documentclass[amsmath,amssymb,aps,twocolumn,prb]{revtex4-2}
\usepackage{graphicx}
\usepackage{dcolumn}
\usepackage{bm}

\usepackage{xcolor}
\usepackage{braket}
\definecolor{dr}{rgb}{0.8,0,0}
\definecolor{dg}{rgb}{0,0.7,0}
\definecolor{turq}{rgb}{0,0.5,0.7}

\newcommand{\ie}{{\it i.e.}, }
\newcommand{\eg}{{\it e.g.}, }

\usepackage{hyperref}
\usepackage{cleveref}
\crefformat{section}{#2Sec.~#1#3}
\crefformat{subsection}{#2Sec.~#1#3}
\crefformat{figure}{#2Fig.~#1#3}
\crefformat{equation}{#2Eq.~#1#3}
\Crefformat{section}{#2Section~#1#3}
\Crefformat{figure}{#2Figure~#1#3}
\Crefformat{equation}{#2Equation~#1#3}

\begin{document}

\preprint{APS/123-QED}

\title{Cluster Expansion Toward Nonlinear Modeling and Classification}
\author{Adrian Stroth}
\author{Claudia Draxl}
\author{Santiago Rigamonti}
\email{srigamonti@physik.hu-berlin.de}
\affiliation{Physics Department and CSMB, Humboldt-Universit{\" a}t zu Berlin, 12489 Berlin, Germany}

\date{\today}

\begin{abstract}
A quantitative first-principles description of complex substitutional materials like alloys is challenging due to the vast number of configurations and the high computational cost of solving the quantum-mechanical problem. Therefore, materials properties must be modeled. The Cluster Expansion (CE) method is widely used for this purpose, but it struggles with properties that exhibit non-linear dependencies on composition, often failing even in a qualitative description. By looking at CE through the lens of machine learning, we resolve this severe problem and introduce a \textit{non-linear CE} approach, yielding extremely accurate and computationally efficient results as demonstrated by distinct examples. 
\end{abstract}

\maketitle
\section{Introduction}
Modeling the intricate relationship between the atomic structure of a material and its properties, which are governed by quantum mechanics, is a prerequisite for the physical description and understanding of materials. For complex materials, like alloys or similar systems, exhibiting impurities or substitutional atoms, this is a challenging task, owing to a combinatorial explosion of possible configurations. Here, essential tasks such as searching for ground states or computing finite-temperature properties, \eg with Monte Carlo simulations, necessitate a huge number of property evaluations, rendering direct first-principles approaches infeasible. One of the earliest and most successful approaches to model such relationships is the Cluster Expansion (CE) method \cite{Sanchez1984, Connolly1983}. CE allows, in principle, for the exact expansion of any property $P$, depending on the atomic configuration $\bm{\sigma}$, in terms of \textit{cluster functions} $\Gamma_{\bm{\alpha}}(\bm{\sigma})$: 
\begin{align}
P(\bm{\sigma}) = \sum_{\bm{\alpha}} J_{\bm{\alpha}} \Gamma_{\bm{\alpha}}(\bm{\sigma}),\label{eq:ce}
\end{align}
where the expansion coefficients \( J_{\bm{\alpha}} \) for cluster $\bm{\alpha}$, capture the effective many-body interactions between the atomic sites. As a result, CE has become a widely used tool for studying alloying phenomena, including the description of phase diagrams and order-disorder transitions for both bulk and surface systems.
 
Despite many successes, CE has severe limitations when dealing with properties that exhibit a nonlinear dependence on the alloy concentration, as highlighted in Ref.~\cite{Sanchez2010}: In the thermodynamic limit, the expansion cannot be converged with a finite number of terms. Considering that nonlinear dependencies are very common in properties that are typically modeled by CE, such as the energy of formation, it is noteworthy that CE is nevertheless used effectively. A partial justification has been provided in Ref.~\cite{Mueller2017}, where it was found that Eq.\eqref{eq:ce} is able to capture nonlinearities even with a finite number of terms when the system is \textit{fully disordered}. This is the limiting case where the occupancy of a crystal site is independent of the neighboring sites, a situation that typically occurs at very high temperatures, or as a result of quenching from a high-temperature phase, or arising from an intrinsic property of the alloy. 

However, this explanation does not solve the fundamental problem of the general lack of convergence~\cite{Sanchez20170}, which becomes apparent when high accuracy of the model is required and when correlations play an important role. This was strikingly demonstrated, for example, in the case of clathrate alloys~\cite{Troppenz2017,Troppenz2021}, where the stable configurations are highly correlated, such that the occupancy of a crystal site strongly depends on the occupancy of neighboring sites even at high temperatures. In this example, convergence of the CE could only be achieved by splitting the model into concentration ranges, each of which exhibited mostly linear behavior~\cite{Troppenz2017}. 

In this Letter, we propose a straightforward approach that effectively solves the long-standing problem and makes CE useful for the nonlinear modeling of material properties. This approach adds a machine-learning aspect to CE to improve the expressiveness of the models by adding nonlinearities. This paves the way for applying sparsity-promoting compressed-sensing methods in CE and achieving convergence with a finite number of terms. After introducing the formalism, we first apply our method to the Redlich-Kister model of the Gibbs free energy of alloys. This simple example already clearly demonstrates how our novel approach makes CE applicable to powerful ML techniques that prioritize sparsity. We then apply our method to the challenging case of the clathrate alloys mentioned above. We show that our method yields improved models without the need to split the concentration range, for both the energy of mixing and the electronic bandgap. Finally, by fitting a support vector machine, we demonstrate how the method can classify metallic and semiconducting clathrates. Most importantly, we show that our method yields models that are not only more accurate but also computationally cheaper to evaluate than standard CE. This offers enormous advantages for exploring the vast configurational space of substitutional and defected systems. Previous studies have used machine learning to optimize the expansion coefficients $J_{\bm{\alpha}}$ employing methods like Bayesian optimization \cite{Mueller2009} and compressed sensing \cite{Nelson2013}, but without going beyond the standard CE model space of Eq.~(\ref{eq:ce}).

\section{nonlinear cluster expansion}
We start by recasting Eq.~(\ref{eq:ce}) in a form amenable for machine learning. We consider a crystal represented by a supercell, which consists of a periodic repetition of $n_{pl}$ parent lattices. In such a system, in the sum of Eq.~(\ref{eq:ce}), many clusters $\bm{\alpha}$ will be symmetrically equivalent to each other, thus having equal expansion coefficients $J_{\bm{\alpha}}$. For intensive properties, the sum can be then expressed as 
\begin{align}
p(\bm{\sigma}) = \frac{1}{n_{pl}}\sum_{\bm{\alpha}}^{\rm{s.i.}}J_{\bm{\alpha}} \sum_{\beta\equiv\alpha} \Gamma_{\bm{\beta}}(\bm{\sigma})=\sum_{\bm{\alpha}}^{\rm{s.i.}}m_{\bm{\alpha}}J_{\bm{\alpha}}\langle\Gamma_{\bm{\alpha}}(\bm{\sigma})\rangle.\label{eq:ce2}
\end{align}
Here, the sum on the right runs over symmetrically inequivalent (s.i.) clusters, $\equiv$ denotes equivalence, $m_{\bm{\alpha}}$ is the multiplicity of cluster $\bm{\alpha}$  \cite{Walle2009}, \ie the number of clusters equivalent under the point group symmetries of the parent lattice, and $\langle\cdot\rangle$ denotes the average over the $m_{\bm{\alpha}}n_{pl}$ equivalent clusters in the supercell. In the thermodynamic limit $n_{pl}\rightarrow\infty$, the number of s.i. clusters is infinite, so in practice the basis must be cut off. The CE model can then be written as 
\begin{align}
    \hat{p}(\bm{\sigma})=\bm{X}(\bm{\sigma})^\top \bm{{\cal J}}.\label{eq:lince}
\end{align}
Here, $\bm{X}(\bm{\sigma})$ is the vector of cluster correlations with components $X_k(\bm{\sigma})=\langle\Gamma_{\bm{\alpha}_k}(\bm{\sigma})\rangle$; ${\cal J}_k=m_{\bm{\alpha}_k}J_{\bm{\alpha}_k}$; $k=1,2,..., n_c$ is an index to indicate the cluster being considered, and $n_c$ is the number of clusters. This is the standard approach to build models in CE. 

Eq.\eqref{eq:lince} can, however, also be interpreted as a machine-learning (ML) problem. In this language, $X_k(\bm{\sigma})$ are input features, and the equation is linear in them. To go beyond linearity, we make use of a common ML technique \cite{Hastie2001} by creating new variables out of the input $\bm{X}$. We propose augmenting the input space by nonlinear transformations $f_j(\bm{X}):\mathbb{R}^{n_c}\rightarrow \mathbb{R}$, $j=1,2,...,q$, such that the predictions are given by
\begin{align}
    \hat{p}(\bm{\sigma})=\bm{f}\left(\bm{X}(\bm{\sigma})\right)^\top\bm{{\cal K}}.
    \label{eq:nlince}
\end{align}
This is a linear model in the $q$-vector of the coefficients $\bm{{\cal K}}$. The model is also linear in the new $q$ features $f_j$. 
In this work, we choose a polynomial basis of the form \( f_j(\bm{X})=X_1^{n_1}X_2^{n_2} ... \) with the condition \( \sum n_i < d_{\text{max}} \). In addition, one can determine other convenient sets of functions, such as \( f_j(\bm{X}) = \sin X_k \) or \( \exp(X_l/\lVert \bm{X} \rVert) \), using symbolic regression techniques~\cite{Koza1992, Ouyang2018}. The polynomial basis has clear advantages, notably providing a solution to the long-standing \( x^2 \) problem in cluster expansion, as shown in Appendix\ref{app:thex2problem}. Once the features are defined, a model can be constructed by considering a set of $n$ training data points, $\{\bm{\sigma}_1, \bm{\sigma}_2, ... \bm{\sigma}_{n}\}$, for which the property $p_i=p(\bm{\sigma}_i)$ is known, \eg from \textit{ab initio} calculations. The predictions for the training data $\bm{p}^\top=(p_1, p_2, ..., p_n)$ can be written as
\begin{align}
   \hat{\bm{p}}=\bm{F}\bm{{\cal K}},
\end{align}
with the matrix elements of $\bm{F}$ being $F_{ij}=f_j(\bm{X}(\bm{\sigma}_i))$. This approach paves the way for applying techniques of ML and statistical analysis to the CE. This includes compressed-sensing approaches to promote model sparsity and filter noise, support vector machines for classification, and more. 

\begin{figure}[b]
    \centering
    \includegraphics[width=1.0\columnwidth]{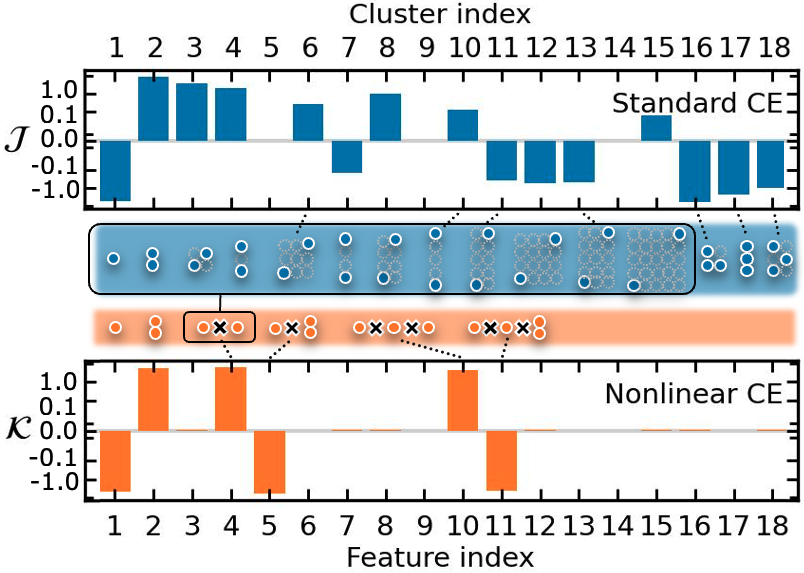}
    \caption{
    Cluster expansion of the Gibbs energy in a second-order Redlich-Kister model. Effective cluster interactions of a standard CE, $\bm{{\cal J}}$  (top) and  coefficients of a nonlinear CE, $\bm{{\cal K}}$ (bottom). Models were selected with LASSO. The middle panel shows schematically the clusters of the standard CE and the nonlinear features of the nonlinear CE (see  Eq.~(\ref{eq:scheme})).\label{fig:toy_model_ecis}}
\end{figure}

\section{Applications}
\subsection{Redlich-Kister model of the Gibbs energy\label{ssec:rkmodel}}
In the following, we demonstrate the success of the nonlinear CE for regression and classification by four distinct examples. The first one is the Redlich-Kister model, where we consider a model Gibbs energy of a binary alloy at zero temperature given by
\begin{align}
    G(\bm{\sigma})=x_A \ \! g_0+x_B  \ \! g_1+\frac{1}{2}J(x_B-x_A)  \ \!x_{AB}(\bm{\sigma}),\label{eq:rkmodel}
\end{align}
with $x_A$, $x_B$, and $x_{AB}$ being the concentrations of species $A$, $B$, and nearest-neighbor (nn) pairs $AB$, respectively. The interaction between neighboring species A and B depends non-linearly on the concentration difference $y=x_B-x_A$ as $J(y)=\sum_{\nu=0}^2\omega_\nu y^\nu$. In the disordered-alloy limit, $x_{AB}(\bm{\sigma})/2\rightarrow x_Ax_B$, and Eq.\eqref{eq:rkmodel} corresponds to a second-order Redlich-Kister expansion of the Gibbs energy \cite{Saunders1998}.

We have generated 132 random structures of a two-dimensional binary alloy with concentrations $x\in[0,1]$ on an $8\times8$ square lattice, and computed the energies using Eq.\eqref{eq:rkmodel} for each of them. The five parameters $g_{0,1}$, $\omega_{0,1,2}$, and the computed energies are provided in the Supporting Information \cite{SI}. The data are then fitted with a standard CE using a cluster basis with up to three-body interactions and a LASSO estimator \cite{Tibshirani1996,Nelson2013}. The latter, based on compressed-sensing and thus aimed at finding sparse models from sparse signals, fails in its mission, as can be seen from the upper panel of Fig.\ref{fig:toy_model_ecis}, which shows a dense model where almost all coefficients are different from zero. In contrast, by performing a nonlinear CE using polynomials up to degree 3 and up to two-body interactions, the application of LASSO yields a sparse model that fits the data exactly, with just six features (lower panel), namely $X_{1b}$, $X_{2nn}$, $X_{1b}^2$, $X_{1b}X_{2nn}$, $X_{1b}^3$, and $X_{1b}^2X_{2nn}$. Here, $X_{1b}$ and $X_{2nn}$ are the one-body and two-body cluster correlations defined, respectively, as $X_{1\rm{b}}(\bm{\sigma})=\langle\sigma_i\rangle/N$ and $X_{2nn}=\langle\sigma_i\sigma_j\rangle/N_{2nn}$, where $\sigma_i=0$ ($1$) if species A (B) occupies crystal site $i$. $N$ and $N_{2nn}$ are the total number of crystal sites and of nn pairs, respectively. Note that the solution found by LASSO matches exactly the generating RK model, as can be obtained analytically by considering the relation between concentrations and cluster correlations, namely $x_B=1-x_A=X_{1b}(\bm{\sigma})$ and $x_{AB}(\bm{\sigma})=2\left[X_{1b}(\bm{\sigma})-X_{2nn}(\bm{\sigma})\right]$, and replacing in Eq.~(\ref{eq:rkmodel})\cite{SI}. We show in Appendix \ref{app:hyperpar} that the result holds in a larger setting, comparing nonlinear and standard CE with clusters up to 6 points and the 14th neighbor.

It is interesting to consider which clusters and features each approach selects. Standard CE includes a large number of 2-body clusters of increasing range, plus 3-body clusters, as shown in the middle panel of Fig. \ref{fig:toy_model_ecis}. In contrast, nonlinear CE selects a few features consisting of nonlinear combinations of compact clusters. These are represented in the middle panel of  Fig. \ref{fig:toy_model_ecis} in an intuitive way, \eg for feature number 11
\begin{align}
    \bullet\!\times\! \bullet\!\times\! \rotatebox{90}{$\!\bullet\bullet\,$}\rightarrow X_{1b}^2X_{2nn} .\label{eq:scheme}
\end{align}
Notably, the nonlinear feature 4, $\bullet\!\times\! \bullet$, contains all the 2-body clusters of the standard CE (indicated with a box in the middle panel of the figure) plus infinite many more, as explained in detail in Sec. \ref{sec:why} (see also Appendix\ref{app:reexpansion}).

\subsection{Energy of mixing of intermetallic clathrate alloys}
Intermetallic clathrate alloys embody the phonon-glass--electron-crystal concept \cite{Slack1997}, being thus promising for thermoelectric energy-conversion applications \cite{Nolas1998,Snyder2008}. The CE modeling of the energy of mixing, as demonstrated with the example of Ba$_8$Al$_{x}$Si$_{46-x}$, ($x\in[6,16]$), turned out to be challenging \cite{Troppenz2017}, since substitution of Si by Al atoms leads to drastic changes of the effective interactions between atomic sites around an Al concentration of $x=12$. As a result, standard CE is unable to cope with this situation, and only a combined model (named "CE$_{\rm{split}}$" below) consisting of two individual standard CE models for concentration ranges up to and above \textbf{$x=12$}, respectively, yielded the required accuracy of around 1~meV/atom. 

Here we demonstrate that a nonlinear CE produces a single model for the entire concentration range, while yielding even better performance than both the standard CE and CE$_{\rm{split}}$ with a similar number of features as cluster functions used in the other schemes. 

\begin{figure}[htb]
    \centering
    \includegraphics[width=\columnwidth]{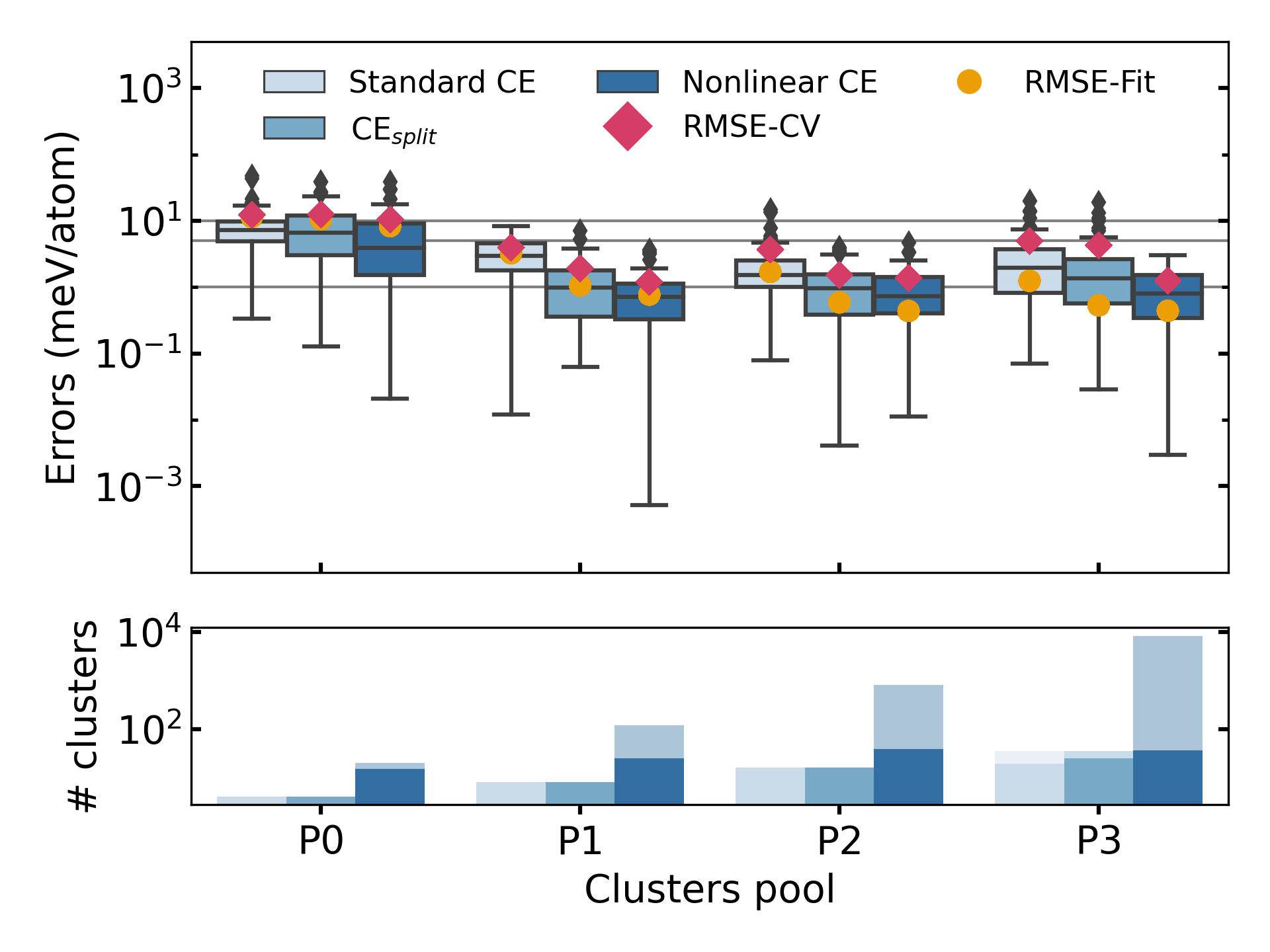}
    \caption{Upper panel: Performance of CE models for the energy of mixing of Ba$_8$Al$_{y}$Si$_{46-y}$, $y\in[6,16]$. The boxplots indicate the distribution of absolute errors in the LOO-CV. Boxes show interquartile ranges (IQR) from the 25th to 75th percentiles with medians as black lines. Whiskers extend to 1.5$\times$IQR. Outliers are represented by black diamonds. Black horizontal lines indicate the values 1 (target accuracy of the model), 5, and 10meV/atom, for visual reference. Bottom panel: Number of cluster functions (standard CE and CE$_{\rm{split}}$) and number of features (nonlinear CE) of the optimal model. Light-colored bars represent the available pool before model selection. \label{fig:cpool_errors}}
\end{figure}

We employ a dataset of 56 structures taken from Ref.\cite{Troppenz2017}. The calculations are performed with the density-functional-theory package \texttt{exciting} \cite{Gulans2014} in the PBE \cite{Perdew2008} paramterization of the generalized gradient approximation for exchange and correlation effects . Computational details are given in Ref. \cite{Troppenz2017}. \Cref{fig:cpool_errors} depicts the distribution of absolute errors as well as the root mean squared error (RMSE) of the fit and of the leave-one-out (LOO) cross validation (CV) of the  models obtained by standard CE, CE$_{\rm{split}}$, and nonlinear CE. Four cluster basis sets are considered: P0 consists of the three one-point clusters representing the Wykoff sites of the clathrate structure \cite{Troppenz2017}; P1 contains P0 plus the nearest-neighbor 2-body clusters; P2 contains P1 plus the second-neighbor two-body clusters; and P3 contains in addition to P2, also the first-neighbor three-body clusters. In all CE models, LASSO-CV \cite{scikit-learn} regression for cluster selection and fitting is used, and the nonlinear features are expanded to the third order.

The figure clearly shows that the nonlinear CE and CE$_{\rm{split}}$ are more accurate than the standard CE for all cases, except for standard CE with P0, which is as accurate as CE$_{\rm{split}}$. For standard CE and CE$_{\rm{split}}$, an optimal RMSE-CV is found with P2. In contrast, the nonlinear CE finds the best model already based on the most simple basis, P1. This indicates that larger clusters do not yield improved models, but including nonlinearities and including nearest-neighbor 2-body correlations is sufficient. Larger basis sets merely lead to overfitting as demonstrated by the increase of the RMSE-CV for P2 and P3 in the nonlinear CE and for P3 in standard CE and CE$_{\rm{split}}$. As expected, the RMSE-Fit decreases with increasing basis-set size. Interestingly, in the nonlinear CE, the number of selected features --indicated by the dark blue bars in the lower panel of Fig.\ref{fig:cpool_errors}-- remains almost constant across all basis sets, despite the rapidly growing number of features going to P2 and P3. The optimal model incorporates only 25 features, characterized by RMSE-Fit = 0.77 meV/atom and RMSE-CV = 1.2 meV/atom. The modified Bayesian CE method from Ref. \cite{Mueller2012} captures nonlinearity to some extent by concentration dependence and, as shown in the Ref. \cite{SI}, resembles a standard CE with nonlinearities in 1-point clusters, making it likely to underperform compared to nonlinear CE.

\subsection{Bandgap of mixing of intermetallic clathrate alloys\label{sec:bandgapregression}}

Different configurations of Ba$_8$Al$_{x}$Si$_{46-x}$ with the same composition $x$ distinguish themselves by exhibiting either metallic or semiconducting behavior \cite{Troppenz2023}. Since efficient thermoelectric behavior is expected in small-gap semiconductors, modeling the dependence of the bandgap on the configuration $\bm{\sigma}$ is of paramount importance. We do so, using a dataset with 78 structures (12 metals and 66 semiconductors) combining DFT-GGA datasets from  Refs.~\cite{Troppenz2017} and \cite{Troppenz2023}. Orthogonal matching pursuit for feature selection \cite{Mallat1993} is employed, generalization errors are estimated by averaging ten rounds of 10-fold CV with random splits. The two basis sets used contain 7 and 368 clusters, respectively: P$_s$ is identical to P1 introduced before, and P$_l$ consists of all clusters with up to three-body interactions within the 54-atoms unit cell. For the nonlinear CE, we use a set of polynomial features up to degree 3 from the elements of $P_s$, termed $P_s^3$. The latter contains 120 features.
\begin{figure}[htb]
    \centering
    \includegraphics[width=\columnwidth]{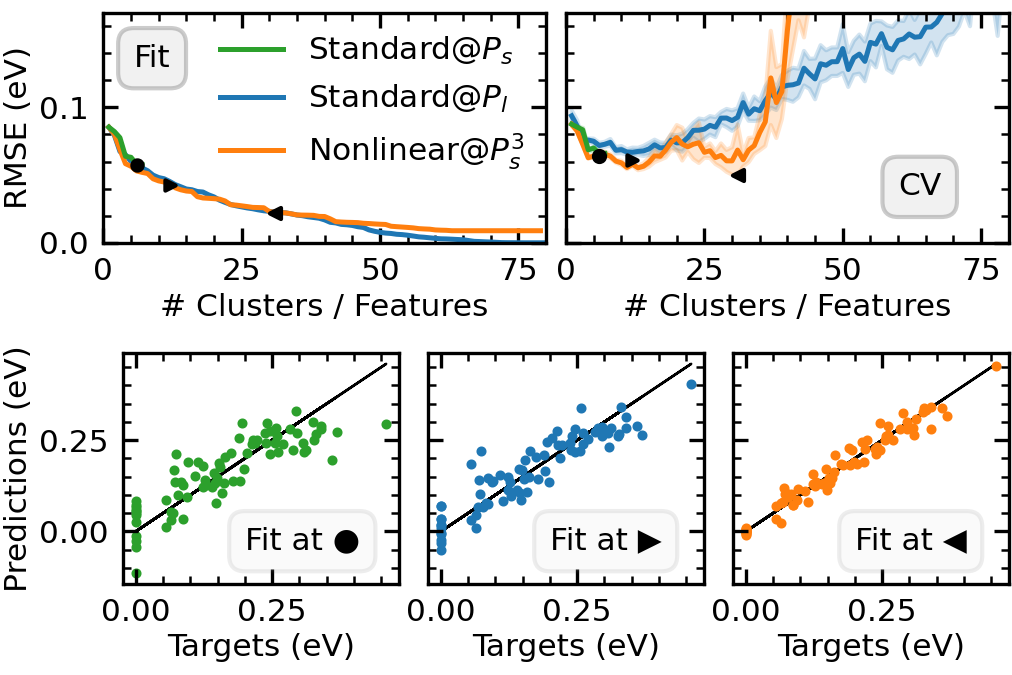}
    \caption{
        Performance of CE models for the bandgap of Ba$_8$Al$_{x}$Si$_{46-x}$ over the whole doping range. Upper panels: RMSE-Fit and RMSE-CV versus the number of clusters for standard CE using $P_s$ and $P_l$, and, respectively, the number of features for nonlinear CE, $P_s^3$. The shaded areas represent the standard deviation (SD) of \mbox{RMSE-CV}. The black dot, right-triangle, and left-triangle indicate the models with lowest \mbox{RMSE-CV}$-$SD for standard CE using $P_s$, standard CE using $P_l$, and nonlinear CE using $P_s^3$ respectively. Bottom panels: Predictions versus target DFT bandgaps for these CE models.
    \label{fig:bandgap}}
\end{figure}

\Cref{fig:bandgap} depicts the performance of the optimal models, \ie those with minimal \mbox{RMSE-CV}. For the standard CE based on the small cluster set $P_s$, the optimal model has 6 clusters, with a \mbox{RMSE-Fit} of 0.057 eV (black circle). Adding many more clusters improves \mbox{RMSE-Fit} slightly (0.043 eV), given by single minimum at 11 clusters (right triangle), very close to the result of the smaller cluster set $P_s$. More complex models overfit, as seen in the quasi-monotonous \mbox{RMSE-CV} increase. Showing a double minimum, the \mbox{RMSE-CV} of the nonlinear CE, based on the $P_s^3$ feature set, exhibits a qualitatively different behavior to the previous cases. The first minimum is located at a number of features similar to the result of the standard CE for the large pool of clusters, but the second minimum at 32 features yields a significantly better model (left triangle), with a \mbox{RMSE-Fit} of 0.022 eV only. Note that the improved generalizability of the nonlinear model is obtained by a nonlinear augmentation of only seven cluster functions. As expected, \mbox{RMSE-Fit} decreases monotonously with model complexity for all three approaches. The standard CE based on the large pool fits the data perfectly at around 75 clusters, while nonlinear CE can achieve an almost perfect fit at around 60 features, based on just 7 clusters. The lower three panels show the accuracy of the predictions. The standard CE shows similar results for both pools of clusters. Notably, both models are inaccurate for metals as well as for the ground-state configuration with the largest bandgap (see \cite{Troppenz2017}). The quality of the predictions of the nonlinear CE is overall significantly better. Most important, it accurately predicts the bandgap of the ground state and finds zero bandgaps for the metals.

\subsection{Nonlinear CE for classification}
This compelling result motivates the use of nonlinear CE for classification tasks, in this case to distinguish metals from semiconductors. To do this, we use the nonlinear CE basis augmentation as input to a support vector machine (SVM) \cite{Cortes1995}. SVMs find a hyperplane that best separates data into different classes. The hyperplane is determined to maximize the margin, which is the largest distance between the nearest data points (support vectors) in different classes. New data points can be classified into one class or another, depending on which side of the hyperplane they fall. We proceed as follows: starting with the pool P1, we consider three transformations of the input, namely the identity (denoted D1) and nonlinear polynomial augmentations of second (D2) and third order (D3). The corresponding number of features is 7, 35, and 119, respectively.

\begin{figure}[htb]
    \centering
    \includegraphics[width=\columnwidth]{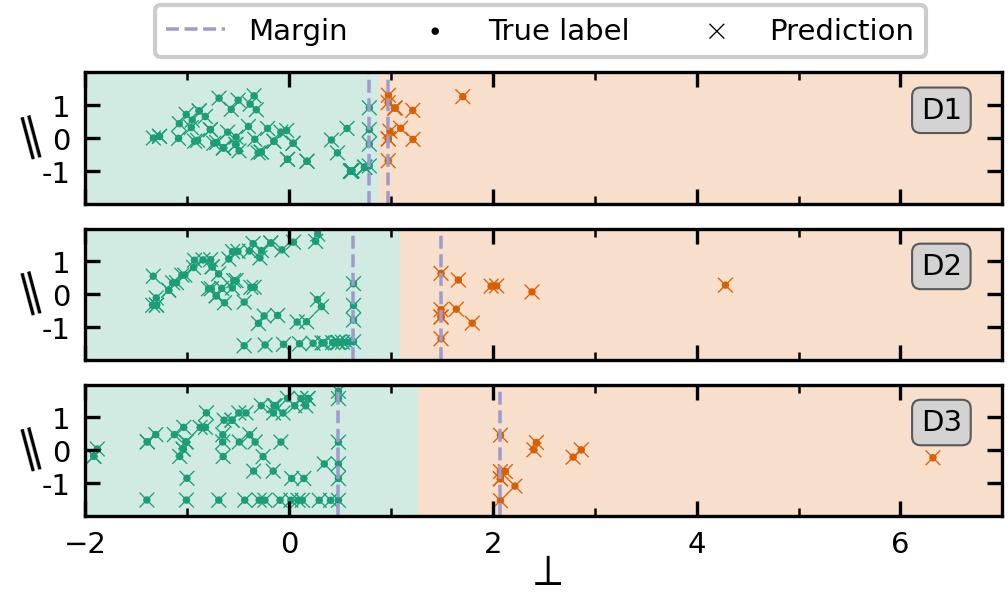}
    \caption{Nonlinear CE-based support vector machine for metal (orange) / semiconductor (turqoise) classification of Ba$_8$Al$_{x}$Si$_{46-x}$. High-dimensional input data are represented in a two-dimensional projection along directions perpendicular ($\perp$) and parallel ($\parallel$) to the decision boundary, which is indicated by the boundary between the two colors. Data separation for classification with original input (D1), and nonlinear polynomial augmentations of second (D2) and third (D3) order.\label{fig:classification}}
\end{figure}

\Cref{fig:classification} shows the resulting classifiers. The obtained linear decision boundary excellently separates metals (orange) from semiconductors (turqoise). To visualize the high-dimensional input D1-3, we map the data points into two-dimensional hyperplanes defined by directions perpendicular and parallel to the decision boundary. Even for the smallest dataset, perfect classification is found. This is indicated by the predictions (crosses) lying on top of the true labels (dots). For the second- and third-order polynomial augmentations, D2 and D3, the margin (thin gray lines) increases. Thus, data separation is notably improved by the nonlinear augmentation. This shows the effectiveness of the nonlinear CE for classification.

\section{Why nonlinear CE works\label{sec:why}}
Our work proposes a change of basis---specifically, a change to an overcomplete basis (see Apendix \ref{app:reexpansion}). While this may not offer an obvious advantage from a purely mathematical perspective, it is indeed crucial for achieving accurate models in cluster expansion from a practical standpoint, and a prerequisite for the successful application of machine learning methods to CE. These two aspects of the basis transformation, \ie its role in ensuring accuracy and its necessity for ML applications, are primarily influenced by (a) the necessity of truncating the basis and (b) the presence of noise in the data or the scarcity of data.

Regarding point (a), note that a very simple property like $x^2$ (the squared concentration) needs a standard CE with an infinite number of terms to be fitted namely, all 2-body clusters up to infinite range\cite{Sanchez2010,Mueller2017,Sanchez20170}, while it can be fitted exactly with just three terms of a nonlinear CE using the polynomial basis (see Appendix \ref{app:thex2problem}). Since in practice, one must cut off the basis, this means that standard CE cannot be converged for such a simple problem. Even worse, cutting off the basis introduces spurious interactions in the fitted models, corresponding to the effect of the clusters left out in the expansion.

Regarding point (b), one must not forget that CE is in essence a data analytics problem, which means fitting to real data. However, the standard CE requires a very large number of terms to accurately describe nonlinear behavior (see previous point), which inherently drives the fitting process toward overfitting. In the presence of noise, the numerous terms in the expansion tend to capture and fit the noise in the data. When data is scarce, the large number of terms results in a strongly underdetermined problem. In both cases, the outcome is the same: the model loses its generalizability. To fix this, one may try adding regularization, which ends up yielding models with few terms that are inaccurate---the case of the band gap of the clathrate in Section \ref{sec:bandgapregression} highlights this clearly, where the standard CE needs a stronger regularization as the nonlinear CE to reduce overfitting. 

In this particular example, the inclusion of 3-body clusters ($P_l$) quickly leads to overfitting and does not yield any improvement with respect to models with 2-body clusters ($P_s$) (see Fig. \ref{fig:bandgap}). When modeling the energy of mixing, overfitting is evident from the fact that optimal performance in the standard CE and CE split models is achieved for \( P2 \), while \( P3 \), which includes 3-body clusters, overfits, as shown in Fig. \ref{fig:cpool_errors}. For the CE modeling of the RK Gibbs energy, the failure of standard CE due to overfitting is dramatic as shown in Appendix \ref{app:hyperpar}.

In contrast, the key factor behind the success of nonlinear CE, lies in its ability to effectively include an infinite number of clusters of a certain order within a single term, in the particular case of polynomial augmentation. This allows nonlinear CE to achieve infinite \textit{range} (maximum distance between any 2 points in a cluster) for specific features. For instance, the term \( X_1^2 \) corresponds to the summation of an infinite number of two-point clusters. This becomes evident by explicitly writing the term using site cluster functions. For the particular case of a simple binary material with a basis $\gamma_0(\sigma_i)=1$ an $\gamma_1(\sigma_i)=\sigma_i$, the expression reads
\begin{align}
    X_1(\bm{\sigma})^2 = \left(\frac{1}{N}\sum_{i=1}^N\sigma_i\right)\left(\frac{1}{N}\sum_{j=1}^N\sigma_j\right)=\frac{1}{N^2}\sum_{i,k=1}^{N}\sigma_i\sigma_j.\label{eq:1pointcorr}
\end{align}
In the thermodynamic limit $N\rightarrow\infty$, the right hand side contains the sum of 2-point correlations of all ranges, \ie
\begin{align}
    N X_1(\bm{\sigma})^2 \rightarrow X_1(\bm{\sigma})+\sum_{r=1}^{\infty}m_{2,r}X_{2,r}(\bm{\sigma}),
\end{align}
where the subindex "$2,r$" means the $r$'th-neighbor 2-point cluster, and the $m's$ are the corresponding cluster multiplicities. A generalization of this relation to arbitrary polynomial terms is given in Appendix \ref{app:reexpansion}.

One may wonder, then, why nonlinear CE, which effectively includes an infinite number of clusters, works, while standard CE fails when more and more clusters are included.

This can be understood easily by comparing the CEs of $x^2$ in the two approaches: In nonlinear CE, using the basis given before Eq.~(\ref{eq:1pointcorr}), the expansion reads
\begin{align}
    x^2={\cal K}X_{1p}^2,\label{eq:cenlinx2}
\end{align} 
involving a single expansion coefficient, ${\cal K}=1$, that can be fitted robustly. Instead, in standard CE, the exact expansion has the form
\begin{align}
     x^2 = J_1X_1(\bm{\sigma})+\sum_{r=1}^{\infty}J_{2,r}m_{2,r}X_{2,r}(\bm{\sigma}),\label{eq:cestdx2}
\end{align}
involving the inclusion of an infinite number of independent terms $J_{2,r}$ and  $J_1$, which for the exact expansion take the values $J_{2,r}=1/N$ and $J_1=1/N$. In noisy or scarce data, fitting such large number of independent terms quickly leads to overfitting, and the necessary cutoff of the range to some $r_{max}$'th neighbor leads to a spurious configuration dependence due to the missing terms in the series.

Finally, since truncated standard cluster expansions have been shown to capture nonlinearities arising specifically from the fully disordered limit (random alloy) (see \cite{Mueller2017}), it is essential to distinguish such nonlinearities from those that are inherent to the property being modeled. To clarify, consider the 2-point nearest-neighbor (nn) correlation. In the disordered limit, this quantity assumes a value of \(x^2\) (see Sec. IC of \cite{SI}). However, this result should not lead one to conclude that an nn 2-point cluster is sufficient to represent an intrinsic \(x^2\) property, \ie an $x^2$ dependence that is present across arbitrary alloy configurations (see \cite{Sanchez20170}), not just in the fully disordered limit. A standard CE capable of fitting sucn an intrinsic \(x^2\) property requires 2-point clusters extending over an infinite range. The fully disordered limit, by the way, neglects the extremely important short-range ordering, which is essential for the description of phase diagrams.

Intrinsic nonlinearities are inherently present in all material properties. For example, a shift in the Fermi level due to changes in the doping concentration leads to a redistribution of the electronic density, such that site energies and inter-atomic interactions change in non-trivial ways. Such effects have been approximately modeled as concentration-dependent \(n\)-point interactions in the past, as in the RK model of our example in Sec.\ref{ssec:rkmodel}. In this context, the nonlinear CE with a polynomial basis that we propose enables to capture nonlinear behavior in a simple and systematic way, leading to converging CEs with significantly fewer terms.

\section{Conclusions}
To summarize, in this Letter, we have introduced a highly accurate yet numerically efficient method, the {\it nonlinear CE}, that solves the longstanding problem of modeling nonlinear properties in the cluster-expansion technique and renders CE amenable to machine learning. We have demonstrated the potential of our approach with various examples. First, we have shown that a Redlich-Kister model of the alloy Gibbs energy can be exactly represented by a nonlinear CE with third-order polynomial features, while standard CE yields a dense, non-converging, representation. This clearly illustrates how our method paves the way for applying sparsity-promoting techniques in CE, such as compressed sensing, which are otherwise ineffective when the standard CE basis is used. Second, the modeling of materials properties was demonstrated with the example of the intermetallic clathrate Ba$_8$Al$_{x}$Si$_{46-x}$. A third-order expansion of a small pool of only 7 clusters, containing first-neighbor two-point interactions, resulted in higher precision and efficiency than a standard CE, for modeling two distinctive properties, namely the energy of mixing and the bandgap. The latter task was particularly challenging, since this material exhibits a metal-to-semiconductor transition. For this reason, standard CE models with up to 368 clusters turned out inaccurate. Finally, we have demonstrated for the same material system that nonlinear CE can also be used for classification tasks, \ie to tell apart metals from semiconductors. We have demonstrated that the addition of nonlinear features further improves the classification accuracy by significantly increasing the margin separating the two classes. Notably, in all considered cases, in addition to showing much better performance, the nonlinear CE models require less data (clusters) than standard CE models. Given that their computation is the most numerically demanding part of property evaluation, our method can significantly speed up or even enable simulations which require large amounts of evaluations, such as finite temperature Monte-Carlo simulations. In short, our method, combining the CE technique with machine learning, promises to become an indispensable tool for the modeling of alloys and similar substitutional systems.

\begin{acknowledgments}
Work partially funded by the Max-Planck Research Network BiGmax. Financial support from the European Union’s Horizon 2020 research and innovation program under the grant agreement N° 951786 (NOMAD CoE) is appreciated. Data files can be downloaded from the NOMAD Repository \cite{Draxl2019}: \url{https://dx.doi.org/10.17172/NOMAD/2023.10.24-1} contains calculations for the energies of mixing and bandgaps of clathrates at compositions $x\in[6,16]$, and \url{http://dx.doi.org/10.17172/NOMAD/2019.10.29-1 } provides additional data for the bandgaps at $x=16$. All standard and non-linear CE models were computed with the CE package \texttt{CELL} \cite{Rigamonti2023,Cell2019}. 
\end{acknowledgments}
\appendix
\section{The $x^2$ problem\label{app:thex2problem}}

\begin{figure}[htb]
    \centering
    \includegraphics[width=0.88\columnwidth]{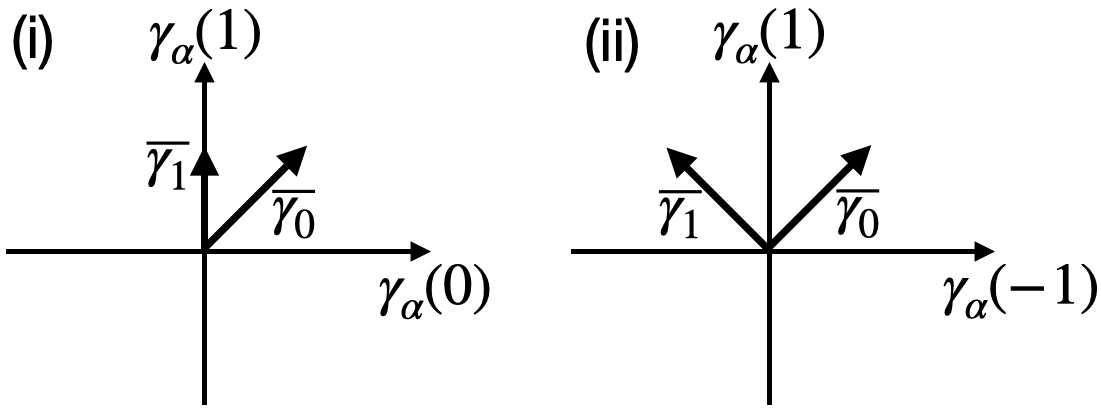}
    \caption{Bases for binaries. (i) Non-orthogonal basis. (ii) Orthogonal basis.\label{fig:basis}}
\end{figure}

In this Appendix, we demonstrate how our approach resolves the "$x^2$ problem" identified in Refs.~\cite{Sanchez2010,Mueller2017,Sanchez20170}. It states that the standard CE requires an infinite number of terms to represent the binary-alloy property \( x^2 \), where \( x \) is the concentration of substituent atoms. We demonstrate this for two bases:  
(i) \( \gamma_0(\sigma_i) = 1 \), \( \gamma_1(\sigma_i) = \sigma_i \), with \( \sigma_i \in \{0,1\} \); and  
(ii) \( \sigma_i \in \{-1,+1\} \), corresponding to the standard domain of Chebyshev polynomials \cite{Walle2009}. A graphical representation of these bases is shown in Fig.\ref{fig:basis}

In basis (i), the nonlinear CE of $x^2$ is that of Eq.\ref{eq:cenlinx2}, involving a single expansion coefficient  ${\cal K}=1$. This expansion follows directly from the observation that $x=\frac{1}{N} \sum_{i=1}^N \sigma_i=X_1(\bm \sigma)$, and consequently
\begin{align}
    x^2=X_1(\bm{\sigma})^2 
\end{align}  
If the more usual cluster basis (ii) is used \cite{Walle2009}, we have 
\begin{align}
    x=\left[X_{\emptyset}+X_{1}(\bm{\sigma})\right]/2,\label{eq:x}
\end{align} 
with $X_{\emptyset}:=1$ and $X_{1}=\frac{1}{N} \sum_{i=1}^N \sigma_i$ being the correlation of the empty and 1-point clusters, respectively. Then, we can write
\begin{equation}
x^2 = \frac{1}{4}X_{\emptyset}+\frac{1}{2}X_{\emptyset}X_{1}(\bm{\sigma})+\frac{1}{4}X_{1}(\bm{\sigma})^2,
\end{equation}
which is an exact expression. It corresponds to a nonlinear CE model with only three features, $f_0(\bm{X})=X_{\emptyset}$, $f_1(\bm{X})=X_{\emptyset}X_{1}$, and $f_2(\bm{X})=X_{1}^2$, and parameters $\bm{{\cal K}}^\top=(\frac{1}{4},\frac{1}{2},\frac{1}{4})$.
\section{Hyperparameter optimization for CE of RK Gibbs energy\label{app:hyperpar}}
We performed extensive hyperparameter optimization using standard CEs with up to 4-point clusters and a range of up to the 14th neighbor, resulting in a pool of 1,281 clusters, and up to 6-point clusters with a range up to the 14th neighbor (10th for the 6-point ones), resulting in a pool of 12,127 clusters. These were compared with calculations using nonlinear CE, based on a cluster pool of up to 2 points, ranging up to the 4th neighbor (just 4 clusters) and polynomial augmentation up to degree 2, 3, and 4, resulting in features spaces of size 20, 55, and 125, respectively.

To obtain optimal models using Lasso regression, we employed an iterative procedure based on multiple train-test splits and model selection. The dataset was randomly partitioned into training (80\%) and test (20\%) subsets, ensuring robust evaluation. We performed 60 independent train-test splits to gather statistical insights on model performance. A logarithmically spaced set of 20 regularization strengths ranging from \(10^{-21}\) to \(10^{1}\) was explored. For each train-test split, the Lasso path was computed, and only coefficients with magnitudes above \(10^{-4}\) were retained. To mitigate instability in the final model selection, a ridge regression with a regularization parameter of \(10^{-10}\) was applied to refine the selected features. The mean squared error on the test set was computed for each resulting model, and errors were aggregated according to the number of retained features. The distribution of test errors across different model sizes was analyzed to determine the optimal complexity of the model. This statistical approach helps on the identification of parsimonious models with minimal prediction error while avoiding overfitting.

\begin{figure}[h]
    \centering
    \includegraphics[width=1.0\columnwidth]{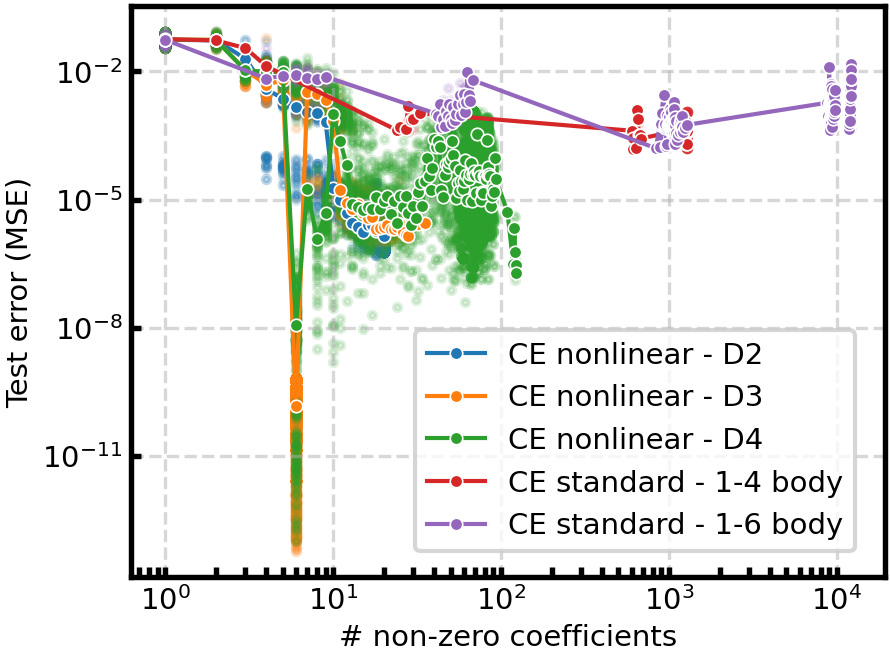}
    \caption{Test error along the LASSO path for 60 independent train-test splits versus the number of non-zero coefficients (model size). Dots represent individual CE models. Solid lines join the median for a given model size. \label{fig:test_errors}}
\end{figure}

Figure \ref{fig:test_errors} shows the results. Standard CEs using cluster pools of 1 to 4 and 1 to 6 clusters are indicated in red and violet, respectively. Nonlinear CEs with polynomial augmentations of degree 2, 3, and 4 are labeled D2, D3, and D4 and indicated in blue, orange, and green, respectively. It is evident that the nonlinear CE is able to find the exact model with just 6 nonzero parameters for degree 3 and degree 4. For larger complexity, all nonlinear CE approaches achieve model accuracy at least two orders of magnitude better than the standard CE. The optimal models for the standard CE with up to 4-body (red) and up to 6-body (violet) clusters have $\sim$1,000 parameters. For the largest model sizes ($\sim$10,000 parameters) obtained with the standard CE with up to 6-body clusters, the models show overfitting. In general, standard CE totally fails to converge to the exact solution and is plagued by overfitting.

\section{Expansion of nonlinear terms in the standard CE basis\label{app:reexpansion}}
In this Appendix, we derive the general structure of the basis augmentation in the polynomial basis expansion of nonlinear CE. The result is that every nonlinear term can be expressed as an infinite sum of basis functions in standard CE, consisting of clusters of limited order (number of points) but up to infinite range (cluster radius).  

A general nonlinear CE polynomial term of degree $d$ reads
\begin{align}
    \prod_{i=1}^k X_{\alpha_i}^{n_i}(\bm{\sigma})=\frac{1}{\prod_{i=1}^k M_{\alpha_i}^{n_i}}\sum_{\overline{S}\in G^d}\prod_{i=1}^d \Gamma_{S_i\alpha_{f(\bm n,i)}}(\bm{\sigma}),\label{eq:expansion1}
\end{align}
with $k\le d$ and $\sum_{i=1}^k n_i = d$, and $M_{\alpha_i}$ the number of inequivalent clusters in the set $\{S_j\alpha_i : S_j\in G\}$, where $S_j$ are the symmetry operations of the symmetry group of the pristine crystal $G$, containing both the point group operations and all translations operations. The lists of symmetry opearations $\overline{S}=(S_1,S_2,...,S_d)$ are taken from the Cartesian product of $G$ with itself $d$ times, denoted $G^d$. The integer function $f(\bm n, i):= \min\{k|i\le \sum_{j=1}^k n_j\}$. 

\begin{figure}[h]
    \centering
    \includegraphics[width=0.9\columnwidth]{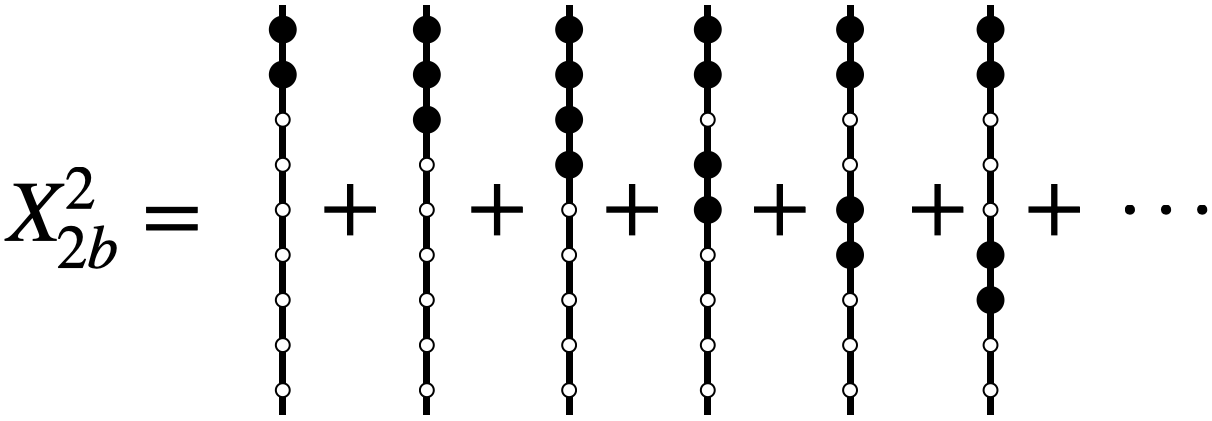}
    \caption{Schematic representation of the expansion of a second order nonlinear CE term in terms of standard CE basis functions.\label{fig:expansion}}
\end{figure}

The product of cluster functions on the right-hand side of Eq.~(\ref{eq:expansion1}) can be re-expressed as a sum of cluster functions of order less or equal $\sum_{i=1}^k n_i \times \text{order} (\alpha_i)$, and the range given by the largest distance between any two points of the aggregated cluster formed by all points in the clusters $\{S_i\alpha_{f(\bm n,i)}|i\in [1,k]\}$. Since the sum in Eq.~(\ref{eq:expansion1}) includes all symmetry operations, single terms of the nonlinear augmentation embody infinite sums of cluster functions up to infinite range. This can be illustrated more simply for an $n$-th order term using the basis (i) defined in Appendix \ref{app:thex2problem}, for which the following relationship can be derived from Eq.~(\ref{eq:expansion1})
\begin{align}
    X_{\alpha}(\sigma)^n = \sum_{\overline{S}\in G^{n-1}} X_{\beta(\overline{S})} (\sigma).\label{eq:expansion2}
\end{align}
Here, the clusters \(\beta(\overline{S})\) take the form 
\begin{align}
    \beta(\overline{S}) = \alpha \oplus \bigoplus_{i=1}^{n-1} (S_i\alpha):=\alpha \lor S_1\alpha \lor \dots \lor S_{n-1}\alpha,
\end{align}
where $\alpha$ is an $N$-vector with 1's and 0's representing the cluster, and $\lor$ represents the component-wise logical "OR" operations. For instance, if $\alpha=(1,1,0,0,0)$ represents a 2-body cluster, and if $S_1\alpha = (0,1,1,0,0)$ and $S_2\alpha = (0,0,0,1,1)$, then $\alpha\oplus S_1\alpha=(1,1,1,0,0)$ and $\alpha\oplus S_2\alpha=(1,1,0,1,1)$ are three- and four-body clusters, respectively. Equation (\ref{eq:expansion2}) implies that the functions \( X_{\alpha}(\sigma)^n \) can be expressed as infinite sums of correlation functions \( X_{\beta}(\sigma) \) of clusters with order $\leq n\times \text{order}(\alpha)$ and up to infinite range. This is represented schematically in Fig.~\ref{fig:expansion} for a second order term from a 2-body cluster (2b).

It is important to note that the polynomial basis is linearly dependent on the standard CE basis only when the complete, infinite basis of the standard CE is considered. As a result, the polynomial augmentation used in the nonlinear CE is overcomplete in this context. However, this overcompleteness is beneficial, because it allows one to express the infinite sums necessary for the modeling of nonlinearities with single terms (which require the evaluation of just one or a few cluster correlations, instead of infinite of them). Apart from this, when a cutoff is imposed on the cluster range---as is standard in all practical applications of CE---\( X_{\alpha}(\sigma)^n \) can no longer be expressed as a linear combination of the correlations in the standard CE basis. This highlights one of the key advantages of the nonlinear CE proposed here: it enables the representation of infinite sums of correlation functions using a finite basis. This is achieved through single nonlinear terms, which are computationally efficient and straightforward to calculate.

\end{document}